\documentclass[twocolumn,showpacs,preprintnumbers,amsmath,amssymb]{revtex4}
\usepackage{amsmath,amssymb,graphicx}
\usepackage{graphicx}
\usepackage{dcolumn}
\usepackage{bm}
\usepackage{epsfig}
\usepackage{here}
\usepackage{float}
\usepackage{amsmath}
\usepackage[usenames]{color}
\usepackage{amsmath}
\usepackage{epsfig}
\usepackage{color}

\newcommand{\bea}{\begin{eqnarray}}
\newcommand{\eea}{\end{eqnarray}}

\begin{document}
\title{Oscillating gravitational potential due to ultralight axion: linear theory}
\author{Jai-chan Hwang${}^{1}$, Hyerim Noh${}^{2}$}
\address{${}^{1}$Center for Theoretical Physics of the Universe,
         Institute for Basic Science (IBS), Daejeon, 34051, Republic of Korea
         \\
         ${}^{2}$Theoretical Astrophysics Group, Korea Astronomy and Space Science Institute, Daejeon, Republic of Korea
         }


\begin{abstract}

We derive oscillating gravitational potential caused by an ultralight axion as the cosmic dark matter, {\it assuming} the Compton wavelength smaller than the horizon scale. A new oscillatory term is found which dominates the previously known one on scales larger than the quantum Jeans-scale. In the context of linear perturbations in cosmology this new term might be more relevant in future observation. Both the quantum stress in density perturbation and the quantum oscillation of the gravitational potential are derived in the zero-shear gauge. We show that the axion fluid in relativistic analysis is quite different from the relativistic and Newtonian zero-pressure fluids, corresponding to Newtonian fluid {\it only} in the density perturbation. The consistency {\it demands} the relativistic analysis valid on scales larger than the Compton wavelength.

\end{abstract}

\maketitle

\section{Introduction}

The presence of a characteristic quantum stress term in a fluid formulation of the Schr\"odinger equation was known immediately after the appearance of the equation by Madelung in 1926 \cite{Madelung-1926}. Ignoring the new term with the quantum origin, which has a role only on a small scale (smaller than the quantum Jeans scale), the fluid is the same as a zero-pressure medium, thus behaving as the cold dark matter. The role of the stress term with extreme small mass was proposed as the fuzzy dark matter in \cite{Hu-Barkana-Gruzinov-2000} and is now an active field of research in the dark matter study with interesting observational consequences, see \cite{FDM-review} for reviews.

In the relativistic perturbation theory, fluid equation with the same quantum stress is derived from the Klein-Gordon equation of a massive scalar field combined with the Einstein equation only assuming the Compton wavelength smaller than the horizon scale \cite{Hwang-Noh-2009, Hwang-Noh-2021}. The result is {\it the same} as in the non-relativistic one; scale larger than the Compton wavelength is demanded for consistency in the relativistic calculation based on the Einstein-Klein-Gordon system, whereas such a condition is not needed in the non-relativistic analysis based on the Schr\"odinger-Poisson system, see below Eq.\ (\ref{relation-ZSG-2}) and \cite{Hwang-Noh-2021}.

Derivation of density perturbation equation known in non-relativistic limit is possible {\it only} in certain gauge condition: the comoving gauge, the zero-shear gauge and the uniform-curvature gauge \cite{Hwang-Noh-2021}. Although the simple non-relativistic equation for density perturbation is derived in three different gauge conditions, {\it only} in the comoving gauge the equation coincides with the zero-pressure fluid system like the cold dark matter. In the other two gauge conditions, the relativistic equations of the zero-pressure fluid are more complicated, and {\it only} in the sub-horizon scale the equations coincide \cite{Hwang-Noh-2021}.

A new quantum effect of the massive scalar field (ultralight axion or fuzzy dark matter) with an astronomically observable consequence was proposed by Khmelnitsky and Rubakov in \cite{Khmelnitsky-Rubakov-2014}. While we ignored the oscillatory terms in deriving the quantum stress by taking the time-average, the new effect is related to the oscillatory gravitational potential caused by the oscillating pressure perturbation with Compton frequency. This is a purely relativistic effect caused by the relativistic pressure, without non-relativistic counterpart.

Here, our aim is deriving the oscillating potential equation in the context of relativistic cosmological perturbation, see Sec.\ \ref{sec:QO}. The situation differs from \cite{Khmelnitsky-Rubakov-2014} which concerns {\it nonlinear} density inhomogeneity in the non-expanding medium. We clarify that the analysis in \cite{Khmelnitsky-Rubakov-2014} is incomplete by not determining the phase of oscillation and ignoring the equation of motion. Here, our study is confined to the linear perturbation theory in cosmology, but considers the full equations in both field and fluid. Determination of the phase leads to a new oscillation term which could be more relevant in cosmological observation. The equation of motion also leads to a condition that the analysis is valid one scales larger than the Compton wavelength. Comparison with \cite{Khmelnitsky-Rubakov-2014} will be made in Sec.\ \ref{sec:discussion}.

In Sec.\ \ref{sec:comparison} we compare equations for the perturbed density, velocity and gravitational potential for three cases: the axion fluid and the zero-pressure fluid both in the ZSG and the Newtonian fluid. The zero-pressure fluid in the ZSG exactly coincides with Newtonian fluid for the perturbed velocity and the gravitational potential, but only in the sub-horizon limit for the perturbed density. Whereas, ignoring the quantum stress term, the axion density perturbation in the ZSG exactly {\it coincides} with the Newtonian one, but both the perturbed velocity and gravitational potential {\it differ} from Newtonian ones. Therefore, for axion, it is convenient to express the result using the density perturbation variable, see Eq.\ (\ref{varphi_chi-eq-axion}).

In Sec.\ \ref{sec:equations} we setup our notation with a summary of the basic equations. In Sec.\ \ref{sec:field-pert} we derive the density and potential perturbation equations with the quantum stress and the quantum oscillation, respectively: these are Eqs.\ (\ref{delta_chi-eq-axion}) and (\ref{varphi_chi-eq-axion}). Section \ref{sec:discussion} is a discussion including comparison with \cite{Khmelnitsky-Rubakov-2014}.

\section{Basic equations}
                                     \label{sec:equations}

We consider a {\it flat} Friedmann cosmology with linear order scalar perturbations. Our metric convention is \cite{Bardeen-1988, Noh-Hwang-2004, Hwang-Noh-2013}
\bea
   & & g_{00}
       = - a^2 \left( 1 + 2 \alpha \right), \quad
       g_{0i} = - a \chi_{,i},
   \nonumber \\
   & &
       g_{ij} = a^2 \left( 1 + 2 \varphi \right)
       \delta_{ij},
   \label{metric}
\eea
where $x^0 = \eta$ with $c dt \equiv a d \eta$, and $a(t)$ is the cosmic scale factor. We imposed a spatial gauge condition which completely removes the spatial gauge degree of freedom; in this sense all perturbation variables can be regarded as spatially gauge-invariant \cite{Bardeen-1988}. The fluid quantities are identified based on the time-like four-vector $u_a$ with $u^a u_a \equiv -1$. In the energy frame, setting the flux four-vector $q_a \equiv 0$, and ignoring the anisotropic stress, we have the energy-momentum tensor \cite{Ellis-1971}
\bea
   & & T_{ab} = \mu u_a u_b
       + p \left( g_{ab} + u_a u_b \right).
   \label{Tab-identification}
\eea
For a massive scalar field, we have
\bea
   & & T_{ab}
       = \phi_{,a} \phi_{,b}
       - {1 \over 2} \left( \phi^{;c} \phi_{,c}
       + {m^2 c^2 \over \hbar^2} \phi^2
       \right) g_{ab}.
   \label{Tab-MSF}
\eea
To the linear order perturbation, we set $\mu \rightarrow \mu + \delta \mu$, $p \rightarrow p + \delta p$, $\phi \rightarrow \phi + \delta \phi$, $u_i \equiv a v_i/c$, and $v_i \equiv - v_{,i}$.

To the background order, we have
\bea
   & & H^2 = {8 \pi G \over 3 c^2} \mu
       + {\Lambda c^2 \over 3}, \quad
       \dot \mu = - 3 H \left( \mu + p \right),
   \label{BG-eqs}
\eea
where $H \equiv \dot a/a$, and $\Lambda$ the cosmological constant. A complete set of perturbation equations is \cite{Bardeen-1988, Noh-Hwang-2004, Hwang-Noh-2013}
\bea
   & & \kappa \equiv 3 H \alpha - 3 \dot \varphi
       - c {\Delta \over a^{2}} \chi,
   \label{eq1} \\
   & & {4 \pi G \over c^2} \delta \mu + H \kappa
       + c^2 {\Delta \over a^{2}} \varphi = 0,
   \label{eq2} \\
   & & \kappa + c {\Delta \over a^{2}} \chi
       - {12 \pi G \over c^4} a \left(
       \mu + p \right) v = 0,
   \label{eq3} \\
   & & \dot \kappa + 2 H \kappa
       + \left( 3 \dot H + c^2 {\Delta \over a^{2}} \right)
       \alpha = {4 \pi G \over c^2}
       \left( \delta \mu
       + 3 \delta p \right),
   \label{eq4} \\
   & & \varphi + \alpha
       - {1 \over c} \left( \dot \chi + H \chi \right)
       = 0,
   \label{eq5} \\
   & & \delta \dot \mu + 3 H \left( \delta \mu
       + \delta p \right)
       = \left( \mu + p \right)
       \left( \kappa - 3 H \alpha + {\Delta \over a} v \right),
   \label{eq6} \\
   & & {1 \over a^{4}} \left[ a^4 \left( \mu
       + p \right) v
       \right]^{\displaystyle\cdot}
       = {c^2 \over a}\left[ \delta p
       + \left( \mu + p \right) \alpha \right].
    \label{eq7}
\eea

For a massive scalar field, the fluid quantities can be identified, using Eqs.\ (\ref{Tab-identification}) and (\ref{Tab-MSF}), as \cite{Hwang-Noh-2021}
\bea
   & & \mu = {1 \over 2 c^2} \left( \dot \phi^2
       + \omega_c^2 \phi^2 \right), \quad
       p = {1 \over 2 c^2} \left( \dot \phi^2
       - \omega_c^2 \phi^2 \right),
   \label{fluid-MSF-BG} \\
   & & \delta \mu
       = {1 \over c^2} \left( \dot \phi \delta \dot \phi
       - \dot \phi^2 \alpha
       + \omega_c^2 \phi \delta \phi \right), \quad
       \left( \mu + p \right) v
       = {1 \over a} \dot \phi \delta \phi,
   \nonumber \\
   & &
       \delta p
       = {1 \over c^2} \left( \dot \phi \delta \dot \phi
       - \dot \phi^2 \alpha
       - \omega_c^2 \phi \delta \phi \right),
   \label{fluid-MSF-pert}
\eea
where $\omega_c \equiv m c^2/\hbar = c/\lambdabar_c$ is the Compton frequency. The equation of motion gives
\bea
   & & \ddot \phi + 3 H \dot \phi + \omega_c^2 \phi
       = 0,
   \label{EOM-BG} \\
   & & \delta \ddot \phi + 3 H \delta \dot \phi
       - c^2 {\Delta \over a^2} \delta \phi
       + \omega_c^2 \delta \phi
   \nonumber \\
   & & \qquad
       = \dot \phi \left( \kappa + \dot \alpha \right)
       + \left( 2 \ddot \phi + 3 H \dot \phi \right) \alpha.
   \label{EOM-pert}
\eea

The above set of perturbation equations is presented without imposing the temporal gauge (hypersurface or slicing)
condition, and all variables used are spatially gauge invariant \cite{Bardeen-1988}. We will mainly consider the zero-shear gauge (ZSG) which sets $\chi \equiv 0$ as the gauge condition \cite{Bardeen-1980, Bardeen-1988}. Under this gauge condition the gauge degrees of freedom are completely fixed, and each variable in this gauge condition has a unique gauge-invariant combination of variables. We will use explicitly gauge-invariant notations
\bea
   \varphi_\chi \equiv \varphi - {1 \over c} H \chi, \quad
       v_\chi \equiv v - {c \over a} \chi
       = - {c \over a} \chi_v, \quad {\it etc.},
   \label{GI}
\eea
where $\varphi_\chi$ is a unique gauge invariant combination which is the same as $\varphi$ in the ZSG setting $\chi \equiv 0$; for gauge transformation properties, see Eq.\ (252) in \cite{Noh-Hwang-2004}.

In the case of field, the fluid quantities constructed from the axion (an oscillating scalar field) qualitatively differ from the ordinary fluid with conventional equation of state. From Eq.\ (\ref{eq4}), using Eqs.\ (\ref{eq1}), (\ref{eq2}) and (\ref{eq5}), we have
\bea
   & & {1 \over a^3} \left[ a^2 \left( a \varphi_{\chi}
       \right)^{\displaystyle{\cdot}} \right]^{\displaystyle{\cdot}}
       = - {4 \pi G \over c^2}
       \left[ \delta p_{\chi}
       + (\mu + p) \alpha_{\chi} \right].
   \label{varphi_chi-p_chi-eq}
\eea
For axion, the right-hand side is pure oscillatory with quantum nature, see Eq.\ (\ref{varphi_chi-eq-axion}). On the other hand, from Eq.\ (\ref{eq1}), using Eqs.\ (\ref{eq2}) and (\ref{eq5}), we have
\bea
   & & 3 H {1 \over a} \left( a \varphi_{\chi}
       \right)^{\displaystyle{\cdot}}
       - c^2 {\Delta \over a^2} \varphi_{\chi}
       = 4 \pi G \delta \varrho_\chi,
   \label{varphi_chi-eq-density}
\eea
which is the $\varphi_{\chi}$ equation supported by the density perturbation; in the {\it sub-horizon} limit, the first term is negligible and we have the Poisson equation in the ZSG.

Now, our task is to derive $\delta p_\chi + (\mu + p) \alpha_\chi$ in Eq.\ (\ref{varphi_chi-p_chi-eq}) in the case of axion as a fuzzy dark matter. We will also show that, for axion, $\delta \varrho_\chi$ satisfies the non-relativistic density perturbation equation known in the fuzzy dark matter. We will show that the relativistic analysis is valid in {\it all} scales bigger than the Compton wavelength \cite{Hwang-Noh-2021}.

\section{Axion perturbation}
                                     \label{sec:field-pert}

Following \cite{Khmelnitsky-Rubakov-2014}, we take an {\it ansatz}
\bea
   & & \phi (t) + \delta \phi ({\bf x}, t)
       = a^{-3/2} [ A + \delta A ({\bf x}, t) ]
   \nonumber \\
   & & \qquad
       \times
       \cos{[\omega_c t + \theta + \delta \theta ({\bf x}, t) ]},
   \label{ansatz}
\eea
with $A$ and $\theta$ constants in time and space.
This is {\it the same} ansatz used in \cite{Hwang-Noh-2009, Hwang-Noh-2021}, and also {\it the same} as the Klein transformation used to derive the Schr\"odinger equation from the Klein-Gordon equation \cite{Klein-1927, Chavanis-Matos-2017}, see Eqs.\ (38) and (54) in \cite{Hwang-Noh-2021}. To the linear order in perturbation, we have
\bea
   & & \phi
       = a^{-3/2} A \cos{(\omega_c t + \theta )},
   \\
   & & \delta \phi
       = a^{-3/2} \left[ \delta A
       \cos{(\omega_c t + \theta)}
       - A \delta \theta \sin{(\omega_c t + \theta)} \right].
\eea

{\it Assuming} $H/\omega_c = \hbar H/(m c^2) = \lambdabar_c / \lambda_H \ll 1$, with $\lambda_H \equiv c/H$ the horizon scale and $\lambdabar_c \equiv \hbar/(mc)$ the Compton wavelength, the fluid quantities in Eqs.\ (\ref{fluid-MSF-BG}) and (\ref{fluid-MSF-pert}) give
\bea
   & & \mu = {\omega_c^2 \over 2 c^2 a^3} A^2, \quad
       p = - \mu \cos{(2 \omega_c t + 2 \theta)},
   \label{fluid-axion-BG}\\
   & & {\delta \mu \over \mu} + (1 + {\rm w}) \alpha
       = 2 {\delta A \over A},
   \nonumber \\
   & & {\delta p \over \mu} + (1 + {\rm w}) \alpha
       = - 2 {\delta A \over A} \cos{(2 \omega_c t + 2 \theta)}
   \nonumber \\
   & & \qquad
       + 2 \delta \theta \sin{(2 \omega_c t + 2 \theta)},
   \nonumber \\
   & & v = {c^2 \over a \omega_c}
       \left( \delta \theta
       - { \sin{(2 \omega_c t + 2 \theta)} \over
       1 - \cos{(2 \omega_c t + 2 \theta)} }
       {\delta A \over A} \right),
   \label{fluid-axion-pert}
\eea
where we set ${\rm w} \equiv p/\mu$. Equation of motion in Eq.\ (\ref{EOM-pert}) gives
\bea
   2 {\delta \dot A \over A}
       - {c^2 \over \omega_c}
       {\Delta \over a^2} \delta \theta
       = \kappa, \quad
       2 \delta \dot \theta
       + {c^2 \over \omega_c} {\Delta \over a^2}
       {\delta A \over A}
       = 2 \omega_c \alpha,
   \label{EOM-axion}
\eea
where we used $\dot \alpha \sim H \alpha \ll \omega_c \alpha$. With the ansatz $A$ and $\theta$ constants, the background equation of motion in Eq.\ (\ref{EOM-BG}) is valid to $(H/\omega_c)^2$ order. 

Now we have the {\it complete} set of equations for axion perturbation without imposing the gauge condition: these are Eqs.\ (\ref{eq1})-(\ref{eq7}), (\ref{fluid-axion-pert}) and (\ref{EOM-axion}).

Before proceeding, we examine $\delta p$-relation in Eq.\ (\ref{fluid-axion-pert}). Without imposing the gauge condition, and without imposing time-average for both $\delta p$, $v$, and $p$, Eq.\ (\ref{fluid-axion-pert}) gives
\bea
   & & {\delta p \over \mu} + (1 + {\rm w}) \alpha
       = \delta + ( 1 + {\rm w} ) \alpha
   \nonumber \\
   & & \qquad
       + 2 {\omega_c \over c^2} a v
       \sin{(2 \omega_c t + 2 \theta)}.
\eea
This is not wrong but apparently the expression misleads the spirit of our ansatz in Eq.\ (\ref{ansatz}) where $\delta A$ and $\delta \theta$ are supposed to be smoothly varying compared with the Compton oscillation. Thus, the right-hand side of this expression should be pure oscillatory; indeed $v$ contains the oscillatory part canceling the first two non-oscillatory terms.

Thus, for a more proper expression, we may take the time-average
(thus ignore the oscillatory part) of the $v$-relation in Eq.\ (\ref{fluid-axion-pert}). In this way, we have $\delta \theta = (\omega_c / c^2) a v$, and
\bea
   & & {\delta p \over \mu} + (1 + {\rm w}) \alpha
       = - 2 {\delta A \over A} \cos{(2 \omega_c t + 2 \theta)}
   \nonumber \\
   & & \qquad
       + 2 {\omega_c \over c^2} a v
       \sin{(2 \omega_c t + 2 \theta)}.
   \label{delta-p}
\eea
In the following, we {\it take} time-average for $v$ and to the background order, thus $p = 0$.

\subsection{Zero-pressure fluid vs. axion fluid}
                                        \label{sec:comparison}

Before proceeding to determine the coefficients of Eq.\ (\ref{delta-p}), thus making Eq.\ (\ref{varphi_chi-p_chi-eq}) a closed form, here we compare the axion fluid in the ZSG with the zero-pressure fluid in the ZSG (thus, relativistic) and the one in Newtonian context. There are subtle differences among these three formulations.

\subsubsection{Axion fluid}

We further {\it take} the time-average for $\delta p$, thus $\delta p/\mu = - \alpha$. For this axion, Eqs.\ (\ref{eq1})-(\ref{eq7}) become
\bea
   & & \kappa \equiv 3 H \alpha - 3 \dot \varphi
       - c {\Delta \over a^{2}} \chi,
   \label{eq1-axion} \\
   & & {4 \pi G \over c^2} \delta \mu + H \kappa
       + c^2 {\Delta \over a^{2}} \varphi = 0,
   \label{eq2-axion} \\
   & & \kappa + c {\Delta \over a^{2}} \chi
       = {12 \pi G \over c^4} a \mu v,
   \label{eq3-axion} \\
   & & \dot \kappa + 2 H \kappa
       + c^2 {\Delta \over a^{2}} \alpha
       = {4 \pi G \over c^2} \delta \mu,
   \label{eq4-axion} \\
   & & \varphi + \alpha
       = {1 \over c} \left( \dot \chi + H \chi \right),
   \label{eq5-axion} \\
   & & \dot \delta
       = \kappa + {\Delta \over a} v,
   \label{eq6-axion} \\
   & & {1 \over a} \left( a v \right)^{\displaystyle\cdot}
       = 0.
    \label{eq7-axion}
\eea
The structure of the equations is quite different compared with the zero-pressure fluid case in Eqs.\ (\ref{eq1})-(\ref{eq7}).

In the ZSG, from Eq.\ (\ref{eq5-axion}) we have $\alpha_\chi = - \varphi_\chi$. From Eqs.\ (\ref{eq2-axion}) and (\ref{eq3-axion}), Eq.\ (\ref{eq7-axion}), Eqs.\ (\ref{eq1-axion}) and (\ref{eq3-axion}), and Eqs.\ (\ref{eq3-axion}) and (\ref{eq6-axion}), respectively, we have
\bea
   & & c^2 {\Delta \over a^2} \varphi_\chi
       + 4 \pi G \varrho \delta_\chi
       = - {12 \pi G \varrho \over c^2} H a v_\chi,
   \label{Poisson-eq-ZSG-axion} \\
   & & \dot v_\chi + H v_\chi = 0,
   \label{dot-v-eq-ZSG-axion} \\
   & & \dot \varphi_\chi + H \varphi_\chi
       = - {4 \pi G \varrho \over c^2} a v_\chi,
   \label{dot-varphi_chi-eq-ZSG-azion} \\
   & & \dot \delta_\chi = {\Delta \over a} v_\chi
       + {12 \pi G \varrho \over c^2} a v_\chi.
   \label{dot-delta-eq-ZSG-axion}
\eea
From these we have
\bea
   & & {1 \over a^3} \left[ a^2 \left( a \varphi_{\chi}
       \right)^{\displaystyle{\cdot}}
       \right]^{\displaystyle{\cdot}} = 0,
   \label{ddot-varphi_chi-eq-axion} \\
   & & \ddot \delta_\chi + 2 H \dot \delta_\chi
       - 4 \pi G \varrho \delta_\chi
       = - c^2 {\Delta \over a^2} \alpha_\chi.
   \label{ddot-delta_chi-eq-axion}
\eea
Equations (\ref{dot-v-eq-ZSG-axion}) and (\ref{ddot-varphi_chi-eq-axion}) are closed form equations for $v_\chi$ and $\varphi_\chi$, and $\alpha_\chi$ in Eq.\ (\ref{ddot-delta_chi-eq-axion}) gives the quantum stress (thus, pure quantum effect) and will be determined in terms of $\delta_\chi$ later, see Eq.\ (\ref{delta_chi-eq-axion}).

\subsubsection{Relativistic zero-pressure fluid}

Now, in the case of zero-pressure fluid, we still have $\alpha_\chi = - \varphi_\chi$, and from Eqs.\ (\ref{eq2}) and (\ref{eq3}), Eq.\ (\ref{eq7}), Eqs.\ (\ref{eq1}) and (\ref{eq3}), and Eqs.\ (\ref{eq3}) and (\ref{eq6}), respectively, we have
\bea
   & & c^2 {\Delta \over a^2} \varphi_\chi
       + 4 \pi G \varrho \delta_\chi
       = - {12 \pi G \varrho \over c^2} H a v_\chi,
   \\
   & & \dot v_\chi + H v_\chi = - {c^2 \over a} \varphi_\chi,
   \label{dot-v-eq-ZSG-fluid} \\
   & & \dot \varphi_\chi + H \varphi_\chi
       = - {4 \pi G \varrho \over c^2} a v_\chi,
   \\
   & & \dot \delta_\chi = {\Delta \over a} v_\chi
       + {12 \pi G \varrho \over c^2} a v_\chi
       + 3 H \varphi_\chi.
   \label{dot-delta-eq-ZSG-fluid}
\eea
Thus, Eqs.\ (\ref{dot-v-eq-ZSG-fluid}) and (\ref{dot-delta-eq-ZSG-fluid}) differ from Eqs. (\ref{dot-v-eq-ZSG-axion}) and (\ref{dot-delta-eq-ZSG-axion}).
From these we have
\bea
   & & {1 \over a^2} \left[ a \left( a v_{\chi}
       \right)^{\displaystyle{\cdot}}
       \right]^{\displaystyle{\cdot}}
       = 4 \pi G \varrho v_\chi,
   \label{ddot-v_chi-eq-fluid} \\
   & & {1 \over a^3} \left[ a^2 \left( a \varphi_{\chi}
       \right)^{\displaystyle{\cdot}}
       \right]^{\displaystyle{\cdot}}
       = 4 \pi G \varrho \varphi_\chi,
   \label{ddot-varphi_chi-eq-fluid} \\
   & & \ddot \delta_{\chi} + 2 H \dot \delta_{\chi}
       - 8 \pi G \varrho \delta_{\chi}
       + {3 H^2 + 6 \dot H + {c^2 \Delta \over a^2} \over
       3 H^2 - \left( 1 + {c^2 \Delta \over 12 \pi G \varrho a^2} \right) {c^2 \Delta \over a^2}}
   \nonumber \\
   & & \qquad \times
       \left[ H \dot \delta_{\chi}
       + \left( 1 + {c^2 \Delta \over 12 \pi G \varrho a^2} \right)
       4 \pi G \varrho \delta_{\chi} \right] = 0.
   \label{ddot-delta_chi-eq-fluid}
\eea
These are closed form equations for the perturbed velocity, potential and density, respectively. {\it All} of these {\it differ} from the corresponding ones for the axion in Eqs.\ (\ref{dot-v-eq-ZSG-axion}), (\ref{ddot-varphi_chi-eq-axion}) and (\ref{ddot-delta_chi-eq-axion}). Only in the sub-horizon limit, Eq.\ (\ref{ddot-delta_chi-eq-fluid}) reproduces Eq.\ (\ref{ddot-delta_chi-eq-axion}) with vanishing $\alpha_\chi$.

\subsubsection{Newtonian fluid}

To the linear order in cosmology context, the Newtonian conservation equations and the Poisson equation give \cite{Hwang-Noh-2021}
\bea
   & & \dot \delta + {1 \over a} \nabla \cdot {\bf v} = 0,
   \\
   & & \dot {\bf v} + H {\bf v} = - {1 \over a} \nabla \Phi,
   \\
   & & {\Delta \over a^2} \Phi = 4 \pi G \varrho \delta.
\eea
We can derive closed form equations as
\bea
   & & {1 \over a^2} \left[ a \left( a \nabla \cdot {\bf v}
       \right)^{\displaystyle{\cdot}}
       \right]^{\displaystyle{\cdot}}
       = 4 \pi G \varrho \nabla \cdot {\bf v},
   \\
   & & {1 \over a^3} \left[ a^2 \left( a \Phi
       \right)^{\displaystyle{\cdot}}
       \right]^{\displaystyle{\cdot}}
       = 4 \pi G \varrho \Phi,
   \\
   & & \ddot \delta + 2 H \dot \delta
       - 4 \pi G \varrho \delta = 0.
   \label{ddot-delta-Newtonian}
\eea

Comparison of the closed form equations in the three formulations reveals that, for a fluid, $\delta_\chi$ in the sub-horizon limit, and $v_\chi$ and $\varphi_\chi$ in all scales have Newtonian correspondence \cite{Hwang-Noh-1999}. In the case of axion, however, ignoring the quantum stress, $\delta_\chi$ in {\it all} scales coincides with the Newtonian $\delta$, whereas both $v_\chi$ and $\varphi_\chi$ {\it differ} from the corresponding Newtonian counterparts in all scales. This implies that for clear understanding in axion, we better express results using $\delta_\chi$ which has exact Newtonian correspondence.

Now, we derive the the quantum stress and quantum oscillation terms in Eqs.\ (\ref{ddot-delta_chi-eq-axion}) and (\ref{varphi_chi-p_chi-eq}) using the complete set of equation for axion in Eqs.\ (\ref{fluid-axion-pert}), (\ref{EOM-axion}) and (\ref{eq1-axion})-(\ref{eq7-axion}).

\subsection{Quantum stress}
                                        \label{sec:QS}

The density perturbation equation for axion fluid was derived for three different gauges (the comoving gauge, the ZSG, and the uniform-curvature gauge) in unified form in \cite{Hwang-Noh-2021}. Here, we present the case in the ZSG, using our different form of ansatz in Eq.\ (\ref{ansatz}).

We have the density perturbation equation in Eq.\ (\ref{ddot-delta_chi-eq-axion}), and to close the equation we need to determine $\alpha_\chi$ in terms of $\delta_\chi$. For that purpose, we can use the fluid quantities in Eq.\ (\ref{fluid-axion-pert}) and the equation of motion in Eq.\ (\ref{EOM-axion}). We take time-average for all fluid variables, and Eq.\ (\ref{fluid-axion-pert}) gives
\bea
   \delta_\chi
       = 2 {\delta A_\chi \over A} - \alpha_\chi, \quad
       {\delta p_\chi \over \mu} = - \alpha_\chi, \quad
       \delta \theta_\chi = {\omega_c \over c^2} a v_\chi.
   \label{relation-ZSG-1}
\eea
The third relation, together with Eq.\ (\ref{eq7-axion}), implies  $\delta \dot \theta_\chi = 0$. Equation (\ref{EOM-axion}) gives
\bea
   & & 2 {\delta \dot A_\chi \over A}
       - {\Delta \over a} v_\chi
       = \kappa_\chi, \quad
       \alpha_\chi
       = {c^2 \over 2 \omega_c^2} {\Delta \over a^2}
       {\delta A_\chi \over A}.
   \label{relation-ZSG-2}
\eea
Equation (\ref{eq6-axion}) gives $\dot \delta_\chi = \kappa_\chi + \Delta v_\chi/a$, and consistency with a combination of the first relations of Eqs.\ (\ref{relation-ZSG-1}) and (\ref{relation-ZSG-2}) {\it demands} $\alpha_\chi \ll \delta_\chi$, thus $(\lambda_c/\lambda)^2 \ll 1$. Thus, we have
\bea
   & & \delta_\chi = 2 {\delta A_\chi \over A}, \quad
       \alpha_\chi = {c^2 \over 4 \omega_c^2}
       {\Delta \over a^2} \delta_\chi.
   \label{relation-ZSG-3}
\eea
Therefore, Eq. (\ref{ddot-delta_chi-eq-axion}) gives
\bea
   & & \ddot \delta_\chi + 2 H \dot \delta_\chi
       - 4 \pi G \varrho \delta_\chi
       = - {\hbar^2 \Delta^2 \over 4 m^2 a^4} \delta_\chi,
   \label{delta_chi-eq-axion}
\eea
where the right-hand side is the quantum stress term; the left-hand side is the well known perturbation equation for a pressureless medium in the comoving gauge and the synchronous gauge \cite{Lifshitz-1946}, and in Newtonian context \cite{Bonnor-1957}.

The competition between gravity and quantum stress terms in Eq.\ (\ref{delta_chi-eq-axion}) gives the quantum Jeans scale
\bea
   & & \lambda_J = {2 \pi a \over k_J}
       = \pi \left( {\hbar^2 \over \pi G \varrho m^2} \right)^{1/4}
       = \sqrt{2 \pi \lambda_c \lambda_H \over \sqrt{6 \Omega_m}}
   \nonumber \\
   & & \qquad
       = {55.6 {\rm kpc} \over \sqrt{ m_{22} \sqrt{\Omega_{m0}}h_{100}}}
       \left( {\varrho_0 \over \varrho} \right)^{1/4},
   \label{Jeans-scale}
\eea
where $m_{22} \equiv m c^2/(10^{-22} {\rm eV})$, $\lambda_c = 0.40 {\rm pc}/m_{22}$, $H_0 = 100 h_{100} {\rm km}/({\rm sec} \cdot {\rm Mpc})$, and $\Omega_m = 8 \pi G \varrho / (3 H^2)$ the density parameter of (axion) dark matter component; subindex $0$ indicates the present epoch.

\subsection{Quantum oscillation}
                                        \label{sec:QO}

In Eq.\ (\ref{varphi_chi-p_chi-eq}), to close the equation, we need $\delta p_{\chi} + (\mu + p) \alpha_{\chi}$ expressed in terms of $\varphi_{\chi}$. To derive the quantum stress we took time-average of $\delta p$ in Eq.\ (\ref{fluid-axion-pert}). Now, by {\it not} taking the time-average of $\delta p_\chi$, we can derive a closed equation for Eq.\ (\ref{varphi_chi-p_chi-eq}).

In the ZSG, Eq.\ (\ref{delta-p}) gives
\bea
   & & {\delta p_\chi \over \mu} + \alpha_\chi
       = - 2 {\delta A_\chi \over A}
       \cos{(2 \omega_c t + 2 \theta)}
   \nonumber \\
   & & \qquad
       + 2 \delta \theta_{\chi}
       \sin{(2 \omega_c t + 2 \theta)}.
   \label{delta-p-ZSG}
\eea
Our task is to express $\delta A_\chi$ and $\delta \theta_\chi$ in terms of perturbation variables like $\varphi_\chi$, $v_\chi$ or $\delta_\chi$. As these coefficient variables are slowly varying compared with the Compton oscillation, in order to determine these now we can freely take the time-average of the pressure perturbation to evaluate them. The relations were already determined in Eqs.\ (\ref{relation-ZSG-1})-(\ref{relation-ZSG-3}).
Together with Eqs.\ (\ref{Poisson-eq-ZSG-axion}) and (\ref{dot-varphi_chi-eq-ZSG-azion}), we have
\bea
   & & - 2 {\delta A_\chi \over A}
       = - \delta_\chi
       = {1 \over 4 \pi G \varrho}
       \left[ c^2 {\Delta \over a^2} \varphi_\chi
       - 3 H {1 \over a} \left( a \varphi_\chi
       \right)^{\displaystyle{\cdot}} \right],
   \nonumber \\
   & & 2 \delta \theta_\chi
       = 2 {\omega_c \over c^2} a v_\chi
       = - {\omega_c \over 2 \pi G \varrho}
        {1 \over a} \left( a \varphi_\chi
       \right)^{\displaystyle{\cdot}}
       = {2 \omega_c a^2 \over c^2 \Delta} \dot \delta_\chi,
   \label{expressions-ZSG}
\eea
where we used sub-horizon limit in the last step. Therefore, we have the relation we are looking for
\bea
   & & {\delta p_\chi \over \mu} + \alpha_\chi
       = {1 \over 4 \pi G \varrho}
       c^2 {\Delta \over a^2} \varphi_\chi
       \cos{(2 \omega_c t + 2 \theta)}
   \nonumber \\
   & & \qquad
       - {\omega_c \over 2 \pi G \varrho}
       {1 \over a} \left( a \varphi_\chi
       \right)^{\displaystyle{\cdot}}
       \sin{(2 \omega_c t + 2 \theta)},
   \label{QO}
\eea
where we ignored $H/\omega_c$ order term. This closes Eq.\ (\ref{varphi_chi-p_chi-eq}).

The perturbation variable $\varphi_\chi$ can be expressed in terms of $\delta_\chi$ and $v_\chi$ using Eq.\ (\ref{expressions-ZSG}). In the case of axion, as {\it only} $\delta_\chi$ has Newtonian correspondence in sub-horizon scale [see below Eq.\ (\ref{ddot-delta-Newtonian})], we use $\delta_\chi$. Equation (\ref{varphi_chi-p_chi-eq}) gives
\bea
   & & {1 \over a^3} \left[ a^2 \left( a \varphi_{\chi}
       \right)^{\displaystyle{\cdot}} \right]^{\displaystyle{\cdot}}
       = 4 \pi G \varrho \Big[ \delta_\chi
       \cos{(2 \omega_c t + 2 \theta)}
   \nonumber \\
   & & \qquad
       + 2 {\omega_c a^2 \over c^2 \Delta} \dot \delta_\chi
       \sin{(2 \omega_c t + 2 \theta)} \Big],
   \label{varphi_chi-eq-axion}
\eea
where we {\it assumed} the sub-horizon limit.

Now, we evaluate the oscillation strength. {\it Assuming} near stationary medium [i.e., ignoring the time dependence of $\varrho$, $a$ and $\delta_\chi$, but with $\dot \delta_\chi \sim \delta_\chi/t_g \sim \sqrt{G \varrho} \delta_\chi$], we have the solution
\bea
   & & \varphi_\chi
       \sim - {4 \pi G \varrho a^2 \over c^2 \Delta} \delta_\chi \bigg[ 1 - {\lambda_c^2 \over 4 \lambda^2}
       \cos{(2 \omega_c t + 2 \theta)}
   \nonumber \\
   & & \qquad
       - {1 \over 8 \sqrt{\pi}} {\lambda_c^2 \over \lambda_J^2}
       \sin{(2 \omega_c t + 2 \theta)} \bigg],
   \label{varphi_chi-solution}
\eea
where the non-oscillatory solution is the one supported by the density perturbation. The two oscillatory solutions are competing with the quantum Jeans scale as the criterion: the sine term dominates over the cosine term for $\lambda > \lambda_J$.

We may set Eq.\ (\ref{varphi_chi-solution}) as \cite{Khmelnitsky-Rubakov-2014}
\bea
   & & \Psi ({\bf x}, t) \equiv - c^2 \varphi_\chi
       = \Psi_d ({\bf x})
       + \Psi_c ({\bf x}) \cos{(2 \omega_c t + 2 \theta)}
   \nonumber \\
   & & \qquad
       + \Psi_s ({\bf x}) \sin{(2 \omega_c t + 2 \theta)},
   \label{Psi-solution}
\eea
where
\bea
   & & \Psi_d \equiv - c^2 \varphi_\chi
       = {4 \pi G \varrho a^2 \over \Delta} \delta_\chi
       = - {1 \over \pi} G \varrho \delta_\chi \lambda^2,
   \nonumber \\
   & & \Psi_c \equiv - \Psi_d {\lambda_c^2 \over 4 \lambda^2}
       = {1 \over 4 \pi} G \varrho \delta_\chi \lambda_c^2,
   \nonumber \\
   & & \Psi_s \sim - {\Psi_d \over 8 \sqrt{\pi}}
       {\lambda_c^2 \over \lambda_J^2}
       \sim {\Psi_c \over 2 \sqrt{\pi}}
       {\lambda^2 \over \lambda_J^2}
       \sim {1 \over 8 \pi \sqrt{\pi}} G \varrho \delta_\chi
       {\lambda^2 \lambda_c^2 \over \lambda_J^2}.
\eea
{\it Assuming} nearly scale invariant $\Psi_d$ in cosmic scale, $\Psi_c$ increases as the size becomes smaller with $\Psi_c \propto 1/\lambda^2$, whereas $\Psi_s$ stays constant independent of the size; $\Psi_s$ is dominant over $\Psi_c$ on $\lambda > \lambda_J$. Amplitudes of the oscillatory terms and the quantum Jeans scale increase as the mass decreases, with $\Psi_c \propto 1/m^2$, $\Psi_s \propto 1/m$ and $\lambda_J \propto 1/\sqrt{m}$. The oscillatory terms have a frequency and dimensionless amplitudes
\bea
   & & f = 2 \nu_c 
       = 4.8 \times 10^{-8} m_{22} {\rm Hz},
   \nonumber \\
   & & {\Psi_c \over c^2} \sim 4.0 \times 10^{-17}
       \left( {1 \over m_{22}}
       {100{\rm kpc} \over \lambda_0}
       {a_0 \over a} \right)^2,
   \nonumber \\
   & &
       {\Psi_s \over c^2} \sim 3.7 \times 10^{-17}
       {\sqrt{\Omega_{m0}} h_{100} \over m_{22}}
       \sqrt{\varrho \over \varrho_0},
\eea
where we used $\Psi_d /c^2 \sim 10^{-5}$. The transition from $\Psi_c$ to $\Psi_s$ occurs near the quantum-Jeans scale in Eq.\ (\ref{Jeans-scale}).

Whether this oscillation in the gravitational potential has any future observational prospect besides the pulsar timing array signals proposed in \cite{Khmelnitsky-Rubakov-2014} is yet to be unraveled. Considering the linear nature of our study and the uncertainty of $\Psi_d$ on scales smaller than the quantum Jeans scale, the $\Psi_s$ term which is scale invariant like $\Psi_d$ might be more relevant in cosmology.

\section{Discussion}
                                     \label{sec:discussion}

Our main result is the derivation of the gravitational potential equation with quantum oscillation in Eq.\ (\ref{varphi_chi-eq-axion}). We also derive the density perturbation equation with the quantum stress in Eq.\ (\ref{delta_chi-eq-axion}), both in the ZSG. Compared with previous studies in \cite{Hwang-Noh-2009, Hwang-Noh-2021, Khmelnitsky-Rubakov-2014}, the quantum stress equation is derived using a different notation in Eq.\ (\ref{ansatz}), and the quantum oscillation equation is rigorously derived in the cosmology context. The quantum oscillation of gravitational potential was previously discovered in Minkowski background \cite{Khmelnitsky-Rubakov-2014}. We derive an additional oscillation term that dominates on scales larger than the quantum Jeans scale.

Here, we compare our result with \cite{Khmelnitsky-Rubakov-2014} which motivated our work. The authors considered Minkowski background. They took an ansatz $\widetilde \phi ({\bf x}, t) = \widetilde A ({\bf x}) \cos{( \omega_c t + \widetilde \theta ({\bf x}) )}$ and did not separate the background and perturbation in fluid quantities; tildes indicate quantities including both the background and perturbation, like $\widetilde \phi ({\bf x}, t) = \phi (t) + \delta \phi ({\bf x}, t)$, etc. In the ZSG, they presented
\bea
   & & \widetilde \mu
       = {\omega_c^2 \over 2 c^2} \widetilde A^2, \quad
       \widetilde p
       = - \widetilde \mu
       \cos{( 2 \omega_c t + 2 \widetilde \theta )},
   \label{KR1} \\
   & & \Delta \Psi = 4 \pi G \widetilde \varrho, \quad
       \ddot \Psi = - 4 \pi G \widetilde \mu
       \cos{( 2 \omega_c t + 2 \widetilde \theta )}.
   \label{KR2}
\eea
These can be compared with our Eqs.\ (\ref{relation-ZSG-3}), (\ref{delta-p-ZSG}), (\ref{varphi_chi-eq-density}) and (\ref{varphi_chi-eq-axion}), respectively. Comparison reveals some correspondence, but metric perturbation ($\alpha$ term) is ignored in \cite{Khmelnitsky-Rubakov-2014}. Including the background fluid quantities in Eq.\ (\ref{KR2}) is not correct as the gravitational potential (metric variable) in Eq.\ (\ref{metric}) is pure perturbed order, but it can be justified, {\it if} the background density is negligible compared with the perturbed part as in the galactic halo considered in \cite{Khmelnitsky-Rubakov-2014}. The sine-part oscillation ($\Psi_s$ term) missing in \cite{Khmelnitsky-Rubakov-2014} can be absorbed to $\delta \theta$, but the authors ignored evaluating it.

If one stops here as in \cite{Khmelnitsky-Rubakov-2014}, without analyzing the equation of motion and not specifying the perturbed phase, the ansatz with $\widetilde A({\bf x})$ in \cite{Khmelnitsky-Rubakov-2014} is fine. However, to determine the perturbed part of the phase, $\widetilde \theta ({\bf x})$, thus recovering the sine-part of oscillation in Eq.\ (\ref{varphi_chi-eq-axion}), we have to determine the remaining fluid quantity $v_i$, and to proceed further we need to consider the equation of motion; in the process, the assumption of $A({\bf x})$ may not be guaranteed and decomposing the perturbation and background would be convenient, see our Eqs.\ (\ref{relation-ZSG-1}) and (\ref{relation-ZSG-2}).

Despite these potential shortcomings, as a major difference, \cite{Khmelnitsky-Rubakov-2014} considered dark matter halo as the medium where the density enhancement is huge compared with the background and {\it claimed} validity of Eqs.\ (\ref{KR1}) and (\ref{KR2}) in that highly nonlinear situation in the matter and field. On galactic halo, density enhancement of the dark matter is estimated to be more than $10^4$ times higher than the background density \cite{Khmelnitsky-Rubakov-2014} and even higher in the soliton core predicted in the fuzzy dark matter simulation based on Schr\"odinger-Poisson system \cite{Schive-etal-2014} reaching up to $10^9$ \cite{deMartino-etal-2017}. The subject is {\it beyond} the scope of our present linear perturbation analysis where we consistently consider linear order perturbations in both metric and matter (i.e., axion field).

To handle the nonlinear situation properly, either the post-Newtonian approximation or the weak gravity limit approximation would be needed. In both cases, analysis of the scalar field equation of motion together with the fluid counterpart should be necessary for proper handling with full consistency. This will be pursued in near future.

\section*{Acknowledgments}

We thank Donghui Jeong for useful discussion. H.N.\ was supported by the National Research Foundation (NRF) of Korea funded by the Korean Government (No.\ 2018R1A2B6002466 and No.\ 2021R1F1A1045515). J.H.\ was supported by IBS under the project code, IBS-R018-D1, and by the NRF of Korea (No.\ NRF-2019R1A2C1003031).

%
%


\end{document}